\documentclass[aps,prb,amsmath,reprint,superscriptaddress]{revtex4-1}
\usepackage{amsmath,txfonts,bm,dcolumn,graphicx,physics}
\begin{document}

\title{Imaginary shift in CASPT2 nuclear gradient and derivative coupling theory}
\author{Jae Woo Park}
\email{jaewoopark@cbnu.ac.kr}
\affiliation{Department of Chemistry, Northwestern University, 2145 Sheridan Rd., Evanston, IL 60208, USA.}
\affiliation{Department of Chemistry, Chungbuk National University, Chungdae-ro 1, Cheongju, Chungbuk 28644, Korea}
\author{Rachael Al-Saadon}
\affiliation{Department of Chemistry, Northwestern University, 2145 Sheridan Rd., Evanston, IL 60208, USA.}
\author{Nils E. Strand}
\affiliation{Department of Chemistry, Northwestern University, 2145 Sheridan Rd., Evanston, IL 60208, USA.}
\author{Toru Shiozaki}
\affiliation{Department of Chemistry, Northwestern University, 2145 Sheridan Rd., Evanston, IL 60208, USA.}
\date{\today}

\begin{abstract}
We report the analytical nuclear gradient theory for complete active space second-order perturbation theory (CASPT2) with imaginary shift,
which is commonly used to avoid divergence of the perturbation expression. 
Our formulation is based on the Lagrangian approach and is an extension of the algorithm for CASPT2 nuclear gradients with real shift.
The working equations are derived and implemented into an efficient parallel program. 
Numerical examples are presented for the ground- and excited-state geometries and conical intersections  of a green fluorescent protein model chromophore, $p-$HBDI$^-$.
We also report timing benchmarks with adenine, $p-$HBDI$^-$, and iron porphyrin.
It is demonstrated that the energies and geometries obtained with the imaginary shift improve accuracy at a minor additional cost which is mainly associated with evaluating the effective density matrix elements for the imaginary shift term.
\end{abstract}

\maketitle

\section{Introduction}
In complete active space second-order perturbation (CASPT2) theory,\cite{Andersson1990JPC,Andersson1992JCP,Pulay2011IJQC,Roosbook2016} one has to use some form of denominator shifts
in order to avoid the so-called intruder states.
To see the role of the shifts, let us briefly recapitulate the CASPT2 theory. 
The CASPT2 theory is formulated as a minimization problem of the Hylleraas functional,
\begin{align}
E^{(2)} = \langle \Phi^{(0)} | \hat{T}^\dagger (\hat{H}^{(0)} - E^{(0)} ) \hat{T} |\Phi^{(0)} \rangle 
+ 2 \langle \Phi^{(0)} | \hat{T}^\dagger \hat{H} |\Phi^{(0)} \rangle,
\label{hyll}
\end{align}
where the excitation operator is defined as
\begin{align}
\hat{T} = \sum_\Omega T_\Omega \hat{E}_\Omega.
\end{align}
$\hat{E}_\Omega$ is the standard spin-free excitation operator with $\Omega$ being the excitation manifold.
The stationary point with respect to the amplitude $T_\Omega$ can be found by solving
\begin{align}
\sum_{\Omega'} \langle \Omega | (\hat{H}^{(0)} - E^{(0)} ) |\Omega' \rangle T_{\Omega'} 
 + \langle \Omega | \hat{H} |\Phi^{(0)} \rangle = 0.
\end{align}
where we introduced $|\Omega\rangle = \hat{E}_\Omega |\Phi^{(0)} \rangle $ for brevity.
Suppose $\omega$ be an orthogonal basis in the expansion space $\Omega$ 
that diagonalizes the first term, namely $\langle \omega | (\hat{H}^{(0)} - E^{(0)} ) |\omega' \rangle = \delta_{\omega\omega'}\Delta_{\omega} $ (note, however, that
this basis is not formed in CASPT2 in practice);
then, this equation can be formally solved as
\begin{align}
T_{\omega} =  -\frac{\langle \omega | \hat{H} |\Phi^{(0)} \rangle}{\Delta_{\omega}}, 
\label{gentamp}
\end{align}
which can be used to calculate the second-order energy $E^{(2)}$.

The intruder state problem stems from the fact that $\Delta_{\omega}$ sometimes vanishes, which leads to divergence of the second-order energies.
To regularize this divergence, several schemes have been developed.
The real shift\cite{Roos1995CPL} modifies the denominator by $1/\Delta \to 1/(\Delta + \epsilon)$
and adds the so-called shift correction using the first-order perturbation theory. This results in the 
expression:
\begin{align}
\frac{1}{\Delta}\to \frac{\Delta}{(\Delta + \epsilon)^2} \approx
\left\{ \begin{array}{ll}
\displaystyle
\frac{\Delta}{\epsilon^2} - \frac{2\Delta^2}{\epsilon^3}
& \Delta \ll 1
\\[8pt]
\displaystyle
\frac{1}{\Delta}-\frac{2\epsilon}{\Delta^2}
& \Delta \gg 1
\end{array}
\right.
\label{realaa}
\end{align}
As one can see, when $\Delta$ is small, the contribution is suppressed by $\epsilon$ in the denominator;
when $\Delta$ is large, it approaches $1/\Delta$.
Our previous CASPT2 nuclear gradient works\cite{MacLeod2015JCP,Vlaisavljevich2016JCTC,Park2017JCTC,Park2017JCTC2} have been based on this real shift scheme.

The imaginary shift\cite{Forsberg1997CPL}
replaces the denominator $1/\Delta$ with $\Re [1/(\Delta + i\epsilon)]$.
As pointed out in the original work, this regularization is more attractive than the real shift, especially for excited-state calculations, because
the imaginary shift approach is guaranteed to be singularity free (owing to the fact that there is no pole away from the real axis).
When the shift correction is included, this is equivalent to the following regularization,
\begin{align}
\frac{1}{\Delta}\to \frac{\Delta(\Delta^2 + 2\epsilon^2)}{(\Delta^2 + \epsilon^2)^2} \approx
\left\{ \begin{array}{ll}
\displaystyle
\frac{\Delta}{\epsilon^2} - \frac{3\Delta^3}{\epsilon^4}
& \Delta \ll 1
\\[8pt]
\displaystyle
\frac{1}{\Delta}-\frac{\epsilon^4}{\Delta^5}
& \Delta \gg 1
\end{array}
\right.
\label{imagaa}
\end{align}
It should be noted that, away from the singularity, the imaginary shift scheme 
has an error of the order of $(\epsilon/\Delta)^4$, whereas the real shift scheme
has an error of the order of $(\epsilon/\Delta)$.
This also supports the superiority of the imaginary shift over the real shift. 

The regularization by the imaginary shift can also be justified 
by the comparison to other regularization schemes that are energy dependent.\cite{Evangelista2014JCP,Lee2018JCTC}
For instance, the driven similarity renormalization group (DSRG) of Evangelista and co-workers,\cite{Evangelista2014JCP,Li2015JCTC}
though derived from a completely different perspective, can be considered to be a form of regularization for CASPT2 (at least) at second order.
In DSRG, the divergence is regularized by 
\begin{align}
\frac{1}{\Delta}\to \frac{1-e^{-\Delta^2/\epsilon^2}}{\Delta}
\approx \left\{ \begin{array}{ll}
\displaystyle
\frac{\Delta}{\epsilon^2} - \frac{\Delta^3}{2\epsilon^4} 
& \Delta \ll 1 \\[8pt]
\displaystyle
\frac{1}{\Delta} -\frac{e^{-\Delta^2/\epsilon^2}}{\Delta}
& \Delta \gg 1
\end{array}
\right.
\end{align}
The small $\Delta$ limit is very similar to the imaginary shift,
while the dumping is exponential in the large $\Delta$ limit.
Though there is a small difference, the similarity between the imaginary shift and DSRG provides another evidence for the effectiveness of the imaginary shift scheme.

Motivated by these observations, we extend in this work the CASPT2 nuclear gradient theory to include the imaginary shift.
The theory, algorithm, working equations, and numerical results are presented in the following. All of the computer programs are implemented in the {\sc bagel} program package,
which is publicly available under the GNU General Public License.\cite{bagel,Shiozaki2018WIREs} 

\section{Theoretical Backgrounds}

In this section, we briefly review the CASPT2 theory with the imaginary shift, first introduced in Ref.~\onlinecite{Forsberg1997CPL}.
Hereafter $i$, $j$, $k$, and $l$ label closed orbitals, $r$, $s$, $t$, and $u$ label active orbitals, $a$, $b$, $c$, and $d$ denote virtual orbitals,
and $x$, $y$, $z$, and $w$ label general orbitals, respectively.
$\Omega$ and $\tilde{\Omega}$ are redundant (non-orthogonal) and orthogonal two-electron excitation manifolds.

\subsection{CASPT2-D Energy Evaluation with the Imaginary Shift}

Since the working equations for CASPT2 with imaginary shift are somewhat complicated,
we first start with a simpler form of CASPT2 that
uses the zeroth-order Hamiltonian without off-diagonal couplings, $\hat{H}^{(0)}_D$ (called CASPT2-D).\cite{Andersson1990JPC}
The CASPT2-D perturbative amplitude in the orthogonal basis $\tilde{\Omega}$ is formally
\begin{align}
\mathcal{T}_{\tilde{\Omega}} & = - \frac{\langle \tilde{\Omega} | \hat{H} |\Phi^{(0)} \rangle }{\Delta_{\tilde{\Omega}} + i\epsilon}.
\end{align}
Note that we choose $\tilde{\Omega}$ such that they diagonalize the zeroth-order Hamiltonian in each excitation subspace, with $\Delta_{\tilde{\Omega}}$ being the associated eigenvalues [see discussions above Eq.~\eqref{gentamp}].
The real part of the amplitudes is
\begin{align}
T_{\tilde{\Omega}} = \Re \left ( \mathcal{T}_{\tilde{\Omega}} \right) & = - \frac{\langle \tilde{\Omega} | \hat{H} |\Phi^{(0)} \rangle \Delta_{\tilde{\Omega}}}{\Delta_{\tilde{\Omega}}^2 + \epsilon^2}
\label{tamp0}
\end{align}
which is to be substituted into the Hylleraas functional [Eq.~\eqref{hyll}] to arrive at the CASPT2-D energy expression
with the shift corrections,
\begin{align}
E^{(2)} = - \sum_{\tilde{\Omega}} \frac{\Delta_{\tilde{\Omega}} (\Delta_{\tilde{\Omega}}^2 + 2 \epsilon^2)}{(\Delta_{\tilde{\Omega}}^2 + \epsilon^2)^2} | \langle {\tilde{\Omega}} | \hat{H} |\Phi^{(0)} \rangle |^2.
\label{imagenergy}
\end{align}
It is known\cite{Forsberg1997CPL} that the same perturbative amplitudes can be obtained by variationally minimizing the following functional,
\begin{align}
\langle \Phi^{(1)}| \hat{H}^{(0)}  - E^{(0)} +  \frac{\epsilon^2}{\Delta_{\tilde{\Omega}}}|\Phi^{(1)}\rangle
+ 2\langle \Phi^{(1)} | \hat{H} |\Phi^{(0)} \rangle,
\label{zeromod}
\end{align}
with $|\Phi^{(1)}\rangle = \hat{T} | \Phi^{(0)}\rangle$.
Taking the derivative with respect to $T_{\tilde{\Omega}}$ and setting it equal to zero, one obtains the amplitude equation,
\begin{align}
\langle \tilde{\Omega} | \hat{H}^{(0)}  - E^{(0)} +  \frac{\epsilon^2}{\Delta_{\tilde{\Omega}}} | \tilde{\Omega}\rangle T_{\tilde{\Omega}} + \langle\tilde{\Omega} | \hat{H} |\Phi^{(0)} \rangle = 0, 
\end{align}
from which one recovers Eq.~\eqref{tamp0}. We can further rewrite this equation as follows using $\tilde{T}_{\tilde{\Omega}} = T_{\tilde{\Omega}}/\Delta_{\tilde{\Omega}}$,
\begin{align}
\langle \tilde{\Omega} | \Delta_{\tilde{\Omega}}(\hat{H}^{(0)}  - E^{(0)}) + \epsilon^2 | \tilde{\Omega}\rangle \tilde{T}_{\tilde{\Omega}}
+ \langle\tilde{\Omega} | \hat{H} |\Phi^{(0)} \rangle = 0,
\end{align}
which is explicitly non-singular.
This formulation is amenable to nuclear gradient formulations and will be used later.

To clarify the procedure for obtaining the orthogonal configurations and denominators in the above,
let us take an excitation class $\hat{E}_{ar,bs}$ as an illustrative example ($\{ar,bs\} \in \Omega$).
Applying this operator to the reference configuration generates
excited configurations that are not orthogonal to each other.
The overlap and zeroth-order Hamiltonian matrix elements between these configurations are
\begin{subequations}
\begin{align}
& \langle \Omega_{ar,bs} | \Omega_{ct,du} \rangle = \delta_{ac} \delta_{bd} \mathcal{S}_{rs,tu}, \\
& \langle \Omega_{ar,bs} | \hat{H}^{(0)}- E^{(0)} | \Omega_{ct,du} \rangle \nonumber\\
&\quad = \delta_{ac} \delta_{bd} \left[\mathcal{F}_{rs,tu}  + (f_{aa} + f_{bb} - E^{(0)})\mathcal{S}_{rs,tu}\right] , \\
& \mathcal{S}_{rs,tu} = \Gamma^{(2)}_{rt,su},\\
& \mathcal{F}_{rs,tu} = \sum_{vw} \Gamma^{(3)}_{rt,su,vw} f_{vw},
\end{align}
\end{subequations}
where $\Gamma^{(n)}$ is an $n$-particle reduced density matrix (RDM) of the reference wave functions.
One can find $V^T_{rs}$ that simultaneously satisfies the following conditions, 
\begin{subequations}
\label{orthogonalcond}
\begin{align}
&\sum_{rstu} V^{T}_{rs} \mathcal{S}_{rs,tu} V^{U}_{tu} = \delta_{TU}, \\
&\sum_{rstu} V^{T}_{rs} \mathcal{F}_{rs,tu} V^{U}_{tu} = \delta_{TU} \phi_{T},
\end{align}
\end{subequations}
where $\phi_T$ is the $T$-th eigenvalue of the Fock matrix in this basis.
The denominator in the orthogonal basis for $\tilde{\Omega}_{ab, T}$ ($\{ab,T\}\in \tilde{\Omega}$) then reads
\begin{align}
\Delta_{ab,T} = f_{aa} + f_{bb} + \phi_T - E_0 \label{DabT}
\end{align}
The denominators for other excitation subspaces are similarly defined. 
The substitution of these expressions into Eq.~\eqref{imagenergy} yields the CASPT2-D energy with imaginary shift.

\subsection{CASPT2 Energy Evaluation with Imaginary Shift\label{energysec}}
When the off-diagonal elements are included in the zeroth-order Hamiltonian,
as in the standard CASPT2 method,
the amplitude equation has to be solved iteratively. 
When the imaginary shift is included, the Hylleraas functional
is to be modified according to Eq.~\eqref{zeromod}, and the amplitude equation for $\tilde{T}_{\tilde{\Omega}}$  becomes
\begin{align}
0 &= \langle\tilde{\Omega}| \Delta_{\tilde{\Omega}} (\hat{H}^{(0)}  - E^{(0)}) +  \epsilon^2|\tilde{\Omega}\rangle \tilde{T}_{\tilde{\Omega}} \nonumber \\
& +
\sum_{\tilde{\Omega}^\prime\neq \tilde{\Omega}} \langle \tilde{\Omega} | \hat{H}^{(0)} |\tilde{\Omega}^\prime \rangle \Delta_{\tilde{\Omega}'}\tilde{T}_{\tilde{\Omega}^\prime}
+ \langle \tilde{\Omega} | \hat{H} |\Phi^{(0)} \rangle.
\label{bbb}
\end{align}
Note that this can only be defined in the orthogonal basis $\tilde{\Omega}$, because the
the shift expression is dependent on the denominator $\Delta_{\tilde{\Omega}}$.

In our implementation, we first evaluate the residual vector, $\sigma$, without the shift term,
in the space of the redundant basis $\Omega$,
\begin{align}
\sigma_{\Omega} = \sum_{\Omega'} \langle \Omega | \hat{H}^{(0)} - E^{(0)} | \Omega' \rangle T_{\Omega' }+
\langle \Omega | \hat{H} | \Phi^{(0)} \rangle.
\label{aaa}
\end{align}
The perturbative amplitudes in the redundant [appearing in Eq.~\eqref{aaa}] and orthogonal [appearing in Eq.~\eqref{bbb}] subspaces are related with each other as
\begin{align}
T_{\Omega} = \sum_{\tilde{\Omega}} \tilde{T}_{\tilde{\Omega}}  \Delta_{\tilde{\Omega}} V_{\tilde{\Omega}}^{\Omega}.
\end{align}
$V$ is defined as in Eq.~\eqref{orthogonalcond}. Note that $V$ is block diagonal with respect to the excitation classes and is independent of the virtual indices.
The residual vector in the redundant basis is then projected to the orthogonal subspace;
after adding the imaginary shift contribution, it reads
\begin{align}
\sigma_{\tilde{\Omega}}^\prime = \sum_{\Omega} \sigma_{\Omega} V_{\tilde{\Omega}}^{\Omega} + \epsilon^2 \tilde{T}_{\tilde{\Omega}},\label{ImagContribution}
\end{align}
using which we update the amplitudes in the orthogonal basis as
\begin{align}
\Delta \tilde{T}_{\tilde{\Omega}}& = - \frac{ \sigma_{\tilde{\Omega}}^\prime}{\Delta_{\tilde{\Omega}}^2 + \epsilon^2}.
\end{align}
This procedure is repeated until convergence is achieved.
At convergence, we compute the second-order energy by inserting $T_\Omega$ in the Hylleraas functional, Eq.~\eqref{hyll}.
This procedure computes energies that implicitly include the shift corrections. 

\subsection{Multistate Extensions of the Imaginary Shift}

In the extended multistate CASPT2 theory, XMS-CASPT2\cite{Finley1998CPL,Granovsky2011JCP,Shiozaki2011JCP3,Vlaisavljevich2016JCTC} with the so-called SS-SR contraction scheme,
a correlated basis state ($| \Phi_L^{(1)} \rangle$) for reference state $L$ is generated 
in a similar procedure to that for the state-specific CASPT2 theory above,
\begin{align}
| \Phi_L^{(1)} \rangle = \hat{T}_{L} | \tilde{L} \rangle = | \tilde{\Omega}_L \rangle T_{L,\tilde{\Omega}}.
\end{align}
In the case of the so-called MS-MR contraction, this equation is to be modified; however, in the following,
we omit the working equations for the MS-MR contraction for brevity, though they are derived and implemented into efficient code as well. 
The first-order wave function for physical state $P$ is then formed as a linear combination of correlated basis states, 
\begin{align}
| \Psi_P^{(1)} \rangle = \sum_L |\Phi^{(1)}_L\rangle R_{LP}.
\end{align}
Note that, in XMS-CASPT2, we use the so-called XMS reference states $| \tilde{L} \rangle$ that diagonalize the Fock operator in the model space,
as first proposed for an uncontracted variant, XMCQDPT.\cite{Granovsky2011JCP}

The unitary matrix elements, $R_{LP}$, are determined by the diagonalization of an effective Hamiltonian $H_\mathrm{eff}$ whose elements are
\begin{align}
H_{\mathrm{eff},LL^\prime} & = \langle \tilde{L} | \hat{H} | \tilde{L}^\prime \rangle  + \langle \tilde{L} | \hat{T}_{L}^\dagger \hat{H} | \tilde{L}^\prime \rangle + \delta_{LL^\prime }\langle \tilde{L} | \hat{T}^\dagger_{L} \hat{H} |\tilde{L} \rangle \nonumber \\
& + 
\delta_{LL^\prime } \langle \tilde{L} | \hat{T}^\dagger_{L} ( \hat{H}^{(0)} - E_L^{(0)} ) \hat{T}_{L} | \tilde{L} \rangle.
\end{align}
The shift is included through the perturbative amplitudes.
The XMS-CASPT2 energy expression for the $P$-th state with the imaginary shift is
\begin{align}
E_{P} & = \sum_{MN} \langle \tilde{M} | \hat{H} | \tilde{N} \rangle R_{MP} R_{NP} + \sum_{LN} R_{LP} R_{NP} \langle \tilde{L} | \hat{T}_{L}^\dagger \hat{H} | \tilde{N} \rangle\nonumber\\
&+ \sum_{L} R_{LP}^2 \left(\langle \tilde{L} | \hat{T}^\dagger_{L} \hat{H} |\tilde{L} \rangle  + \langle \tilde{L} | \hat{T}^\dagger_{L} ( \hat{H}^{(0)} - E_L^{(0)} ) \hat{T}_{L} | \tilde{L} \rangle\right).
\end{align}
Unlike XMS-CASPT2 with real shift, the last term that accounts for the shift correction has to be included explicitly.

\section{Nuclear Gradient Theory for CASPT2 with Imaginary Shift}
\subsection{CASPT2 Lagrangian with Imaginary Shift}

The CASPT2 energy with imaginary shift is not stationary with respect to the amplitudes. The CASPT2 part of the Lagrangian is defined as
\begin{align}
\mathcal{L}_{\mathrm{PT2},P} & = E_{P} + \sum_{L,\tilde{\Omega}} \lambda_{L,\tilde{\Omega}} \sigma_{L,\tilde{\Omega}}
+ \sum_{L,\tilde{\Omega}} \Lambda_{L,\tilde{\Omega}} \left[ \Delta_{L,\tilde{\Omega}} - f_{L,\tilde{\Omega}} \right],
\label{lagpt2}
\end{align}
where $f_{L,\tilde{\Omega}}$ is the explicit form of $\Delta_{L,\tilde{\Omega}}$, e.g., the right-hand side of Eq.~\eqref{DabT}.
Note that $\Delta_{L,\tilde{\Omega}}$ is treated as a parameter that is constrained to a particular value as seen in the last term of Eq.~\eqref{lagpt2}.
This allows us to make the Lagrangian linear in the Hamiltonian, which in turn allows for straightforward definition of the relaxed density matrices.

The stationary condition with respect to the perturbation amplitude, the so-called $\lambda$-equation, is obtained by differentiating this Lagrangian with respect to the amplitude $T_{L}$, as
\begin{align}
0 & = \langle \tilde{\Omega}_L | \Delta_{L,\tilde{\Omega}}(\hat{H}^{(0)} - E_L^{(0)}) + \epsilon^2 | \tilde{\Omega}_L \rangle \tilde{\lambda}_{L,\tilde{\Omega}} \nonumber \\
& + \sum_{\tilde{\Omega}_L^\prime\neq \tilde{\Omega}_L} \langle \tilde{\Omega}_L | \hat{H}^{(0)} |\tilde{\Omega}_L^\prime \rangle \Delta_{L,\tilde{\Omega}'}\tilde{\lambda}_{L,\tilde{\Omega}^\prime} +  \sum_{N} R_{NP} \langle \tilde{\Omega}_L | \hat{H} | \tilde{N} \rangle \nonumber \\
& + R_{LP} \left[\langle \tilde{\Omega}_L  | \hat{H} |\tilde{L} \rangle 
+ 2 \langle \tilde{\Omega}_L  | ( \hat{H}^{(0)} - E_L^{(0)} ) \hat{T}_{L} | \tilde{L} \rangle \right]
\label{Lambda_now}
\end{align}
whose last term is not present in the $\lambda$-equation for CASPT2 with the real shift in the previous reports.\cite{Shiozaki2011JCP3,Vlaisavljevich2016JCTC}
This new term arises because, when including the imaginary shift,
we have to use the explicit Hylleraas functional for the diagonal elements of the effective Hamiltonian. 
The difference is compensated later such that the nuclear gradients remain identical in the limit of $\epsilon=0$.
In addition, by taking a derivative of $\mathcal{L}_{\mathrm{PT2},P}$ with respect to $\Delta_{L,\tilde{\Omega}}$ and setting it to zero, one obtains
\begin{align}
\Lambda_{L,\tilde{\Omega}} = \epsilon^2 \tilde{\lambda}_{L,\tilde{\Omega}} \tilde{T}_{L,\tilde{\Omega}}. \label{LambdaOmega}
\end{align}
With these procedures, $\mathcal{L}_{\mathrm{PT2},P}$ is now stationary with respect to all of the parameters, namely $T$, $\lambda$, $\Delta_{L,\tilde{\Omega}}$, and $\Lambda_{L,\tilde{\Omega}}$. 

Next, we consider the conditions associated with constructing the orthogonal basis functions.
The Lagrangian is augmented to account for the fact that the Fock operator is diagonal in the orthogonal basis $\tilde{\Omega}$ [for instance, Eq.~\eqref{orthogonalcond}] as
\begin{align}
\mathcal{L}_{\mathrm{imag},P} & = \mathcal{L}_{\mathrm{PT2},P}
+ \tr \left[ \mathbf{\bar{z}} \left( \mathbf{V}^\dagger \boldsymbol{\mathcal{F}}\mathbf{V} - \boldsymbol{\phi} \right) \right] \nonumber\\
&- \tr \left[ \mathbf{\bar{X}} \left( \mathbf{V}^\dagger \boldsymbol{\mathcal{S}}\mathbf{V} - \mathbf{1} \right) \right].\label{lagimag}
\end{align}
where $\boldsymbol{\phi}$ is a diagonal matrix whose elements are $\phi_T$.
The diagonal elements of $\bar{\mathbf{z}}$ are obtained by the stationary condition for $\mathcal{L}_{\mathrm{imag},P}$ with respect to variation of $\phi_T$.
For example, $\bar{z}_{TT}$ for the configurations $\tilde{\Omega} \in \{ab,T\}$ is
\begin{align}
\bar{z}_{TT} = - \sum_L \sum_{ab} \Lambda_{L,ab,T}.
\end{align}
The remaining elements of $\bar{\mathbf{X}}$ and $\bar{\mathbf{z}}$ are determined by differentiating the Lagrangian with respect to 
the non-redundant set of parameters for $\mathbf{V}$.
We do so by introducing a unitary rotation $\mathbf{W}$, i.e.,
\begin{align}
\mathbf{V} = \mathbf{V}^{0} \mathbf{W}.
\end{align}
Then, the multipliers are
\begin{align}
\bar{z}_{TU} & = -\frac{1}{2} \frac{\bar{Y}_{TU} - \bar{Y}_{UT}}{\phi_T - \phi_U}, \nonumber \\
\bar{X}_{TU} & = \frac{1+\tau_{TU}}{4} \left(\bar{Y}_{TU} + 2 \bar{z}_{TU} \phi_T \right),
\end{align}
where $\tau_{TU}$ permutes the indices $T$ and $U$. $\bar{Y}_{TU}$ is the derivatives of $\mathcal{L}_{\mathrm{imag},P}$
with respect to $\mathbf{W}$, whose explicit expression for the configurations $\tilde{\Omega} \in \{ab,T\}$ is 
\begin{align}
\bar{Y}_{TU} 
& = \epsilon^2 \sum_L \sum_{ab} \tilde{\lambda}_{L,ab,T} \tilde{T}_{L,ab,U} \left(\Delta_{L,ab,T} - \Delta_{L,ab,U}\right).
\end{align}
With these Lagrange multipliers $\lambda$, $\bar{z}$, and $\bar{X}$, $\mathcal{L}_{\mathrm{imag},P}$ is stationary with respect to any variation of $T$, $\Delta$, and $V$.
The terms associated with the use of the orthogonal basis vanish, as expected, when the shift parameter $\epsilon$ is zero or when the real shift is used,
since $\bar{Y}$ becomes zero in these cases (see the Supporting Information).

\subsection{$Z$-vector Equation}

The total Lagrangian, which is to be made stationary with respect to the CI and molecular orbital (MO) coefficients in the CASSCF procedure, reads\cite{Celani2003JCP,Shiozaki2011JCP3}
\begin{align}
\mathcal{L} & = \mathcal{L}_{\mathrm{imag},P} \nonumber \\
& + \frac{1}{2} \tr \left[ \mathbf{Z} \left( \mathbf{A}-\mathbf{A}^\dagger \right) \right] - \frac{1}{2} \tr \left[ \mathbf{X} \left( \mathbf{C}^\dagger \mathbf{SC} - \mathbf{1} \right) \right] \nonumber \\
& + \sum_N W_N \left[ \sum_{I} z_{I,N} \langle I | \hat{H} - E_N^{\mathrm{ref}} | N \rangle - \frac{1}{2} x_N \left( \langle N | N \rangle - 1 \right)  \right] \nonumber \\
& + \sum_i^{\mathrm{closed}}\sum_{j \neq i}^{\mathrm{closed}} z_{ij}^{c} f_{ij} +  \sum_a^{\mathrm{virtual}}\sum_{b \neq a}^{\mathrm{virtual}} z_{ab}^{c} f_{ab}  +\sum_{MN} w_{MN} \langle \tilde{M} | \hat{f} | \tilde{N} \rangle.
\end{align}
Here, $\mathbf{A}$ is an orbital gradient matrix in CASSCF, $\mathbf{S}$ is an overlap integral in the atomic orbital basis, $\mathbf{C}$ is the MO coefficients.
$N$ labels CASSCF states, whose wave function is $|N\rangle$.
$W_N$ is the weight used in the state averaging scheme, and $I$ labels Slater determinants in the active space.
The terms in the second and third lines define the conditions for the CASSCF wave functions.
The remaining terms account for the fact that the Fock matrix in the MO basis is diagonal in the closed and virtual orbital spaces (including the frozen core approximation) and for the condition
associated with the XMS rotations.\cite{Shiozaki2011JCP3}
The multipliers $\mathbf{Z}$, $\mathbf{z}$, and $\mathbf{X}$ can be obtained by solving the so-called $Z$-vector equation.\cite{Celani2003JCP,Shiozaki2011JCP3,Vlaisavljevich2016JCTC}
The source terms for the $Z$-vector equations are the derivatives of $\mathcal{L}_{\mathrm{imag},P}$ with respect to the orbital rotation parameters $\kappa_{xy}$ and the CI coefficients $c_{I,N}$; they are
\begin{align}
Y_{xy} & = \frac{\partial \mathcal{L}_{\mathrm{imag},P}}{\partial \kappa_{xy}}, \\
y_{I,N} & = \frac{\partial \mathcal{L}_{\mathrm{imag},P}}{\partial c_{I,N}}.
\end{align}
To evaluate these terms, it is convenient to rewrite $\mathcal{L}_{\mathrm{imag},P}$ 
in a form that separates 
the terms dependent on the molecular integrals and RDMs,
those that are only dependent on the RDMs,
and those that are independent of the molecular integrals or RDMs.
The rewritten expression is
\begin{align}
\mathcal{L}_{\mathrm{imag},P} & = \tr \left( \mathbf{hd} \right) + \tr \left[ \mathbf{g} \left(\mathbf{d}^{(0),\mathrm{SA}}\right) \mathbf{d}^{(2)} \right] + \sum_{kl} \tr \left( \mathbf{K}^{kl} \mathbf{D}^{lk} \right) \nonumber \\
& + \sum_{L} \sum_{n=1}^{3} \tr \left( \mathbf{e}^{(n)S,LL} \mathbf{\Gamma}^{(n),LL} \right) + 2 \sum_{\tilde{\Omega}} \Lambda_{\tilde{\Omega}} \Delta_{\tilde{\Omega}}- \tr \left( \mathbf{\bar{z}} \boldsymbol{\phi} -  \mathbf{\bar{X}} \right).\label{lagrangian}
\end{align}
Here we use the following notations for molecular integrals 
\begin{subequations}
\begin{align}
\left[\mathbf{g}(\mathbf{d})\right]_{xy} & = \sum_{kl} \left[ (xy|zw) d_{zw} - \frac{1}{4}(xw|zy) (d_{zw} + d_{wz}) \right],\\
\mathbf{K}^{zw}_{xy} & = (xz|yw),
\end{align}
\end{subequations}
and for the RDMs
\begin{subequations}
\begin{align}
&\Gamma^{(1),LL}_{rs} = \langle \tilde{L} | \hat{E}_{rs} | \tilde{L} \rangle, \\
&\Gamma^{(2),LL}_{rs,tu} = \langle \tilde{L} | \hat{E}_{rs,tu} | \tilde{L} \rangle, \\
&\Gamma^{(3),LL}_{rs,tu,vw} = \langle \tilde{L} | \hat{E}_{rs,tu,vw} | \tilde{L} \rangle, \\
&d^{(0),\mathrm{SA}}_{rs} = \sum_{L} W_L \Gamma^{(1),LL}_{rs}.
\end{align}
\end{subequations}
The density-like terms $\mathbf{e}^{(1)S}$ to $\mathbf{e}^{(3)S}$ arise from the overlap of the redundant basis, $\boldsymbol{\mathcal{S}}$, as
\begin{align}
\mathbf{e}^{(n)S,MM} = - \sum_{TU} \bar{X}_{TU} \frac{\partial }{\partial \mathbf{\Gamma}^{(n),MM}} \left( \mathbf{V}^\dagger \boldsymbol{\mathcal{S}} \mathbf{V} \right)_{TU}.
\end{align}	
For example, the contribution from $\tilde{\Omega}\in\left\{ ab,T \right\}$ is
\begin{align}
e^{(2)S,MM}_{rs,tu} = - \sum_{TU} \bar{X}_{TU} V^{T}_{rt,M} V^{U}_{su,M}.\label{ens}
\end{align}
The terms that do not depend on the molecular integrals or RDMs 
contribute to neither $Y_{xy}$ nor $y_{I,N}$.
The total one-electron and two-electron density matrices, $\mathbf{d}$ and $\mathbf{D}$, are
\begin{subequations}
\begin{align}
\mathbf{d} & = \mathbf{d}^{(0)} + \mathbf{d}^{(1)} + \mathbf{d}^{(2)}, \\
\mathbf{D} & = \mathbf{D}^{(0)} + \mathbf{D}^{(1)},
\end{align}
\end{subequations}
where the superscripts denote perturbation order.
The zeroth- and first-order contributions are
\begin{subequations}
\begin{align}
d^{(0)}_{xy} & = \sum_{LN} R_{LP} R_{NP} \langle \tilde{L} | \hat{E}_{xy} | \tilde{N} \rangle , \\
D^{(0)}_{xyzw} & = \sum_{LN} R_{LP} R_{NP} \langle \tilde{L} | \hat{E}_{xyzw} | \tilde{N} \rangle , \\
d^{(1)}_{xy} & = \sum_{LN} R_{LP} R_{NP} \langle \tilde{L} | \hat{T}_{L}^\dagger \hat{E}_{xy} | \tilde{N} \rangle \nonumber \\ & + \sum_L R_{LP}^2 \langle \tilde{L} | \hat{T}^\dagger_{L} \hat{E}_{xy} |\tilde{L} \rangle + \langle \tilde{L} | \hat{\lambda}_{L}^\dagger \hat{E}_{xy} | \tilde{L} \rangle, \\
D^{(1)}_{xyzw} & = \sum_{LN} R_{LP} R_{NP} \langle \tilde{L} | \hat{T}_{L}^\dagger \hat{E}_{xyzw} | \tilde{N} \rangle \nonumber \\ & + \sum_L R_{LP}^2 \langle \tilde{L} | \hat{T}^\dagger_{L} \hat{E}_{xyzw} |\tilde{L} \rangle + \langle \tilde{L} | \hat{\lambda}_{L}^\dagger \hat{E}_{xyzw} | \tilde{L} \rangle.
\end{align}
\end{subequations}
The second-order contributions to the correlated density matrix can be divided into three components,
\begin{align}
\mathbf{d}^{(2)} &= \mathbf{d}^{(2)}_{TT} + \mathbf{d}^{(2)}_{T \lambda}  + \mathbf{d}^{(2)}_{\mathrm{shift}},\label{d2tot}
\end{align}
where $\mathbf{d}^{(2)} = \mathbf{d}^{(2)}_{T \lambda}$ in the CASPT2 nuclear gradient theory for the real shift.\cite{Shiozaki2011JCP3,Vlaisavljevich2016JCTC}
The additional terms for the imaginary shift compensate the difference in the $\lambda$-equation.
The first two terms are
\begin{subequations}
\begin{align}
\bar{d}^{(2)}_{TT,xy} &= \sum_{L} R_{LP}^2 \langle \tilde{L} | \hat{T}^\dagger_{L} \hat{E}_{xy} \hat{T}_{L} | \tilde{L} \rangle,\label{dtdl1}\\
\bar{d}^{(2)}_{ T\lambda,xy} &= \sum_{L}\langle \tilde{L} | \hat{T}^\dagger_{L} \hat{E}_{xy} \hat{\lambda}_{L} | \tilde{L} \rangle,\label{dtdl} \\
d^{(2)}_{TT,xy} &= 
\left\{ \begin{array}{ll}
	\displaystyle
	\bar{d}^{(2)}_{TT,xy} - \sum_L N_L^{TT} \langle \tilde{L} | \hat{E}_{xy} | \tilde{L} \rangle
	& {x,y} \in {r,s} \\[8pt]
	\displaystyle
	\bar{d}^{(2)}_{TT,xy}
	& \mathrm{otherwise}
	\end{array} \right. \\
d^{(2)}_{ T\lambda,xy} &= 
\left\{ \begin{array}{ll}
\displaystyle
\bar{d}^{(2)}_{T\lambda,xy} - \sum_L N_L^{\lambda T} \langle \tilde{L} | \hat{E}_{xy} | \tilde{L} \rangle
& {x,y} \in {r,s} \\[8pt]
\displaystyle
\bar{d}^{(2)}_{T\lambda,xy}
& \mathrm{otherwise} \end{array} \right.
\end{align}
\end{subequations}
in which we used
\begin{subequations}
\begin{align}
&N_L^{TT} = R_{LP}^2 \langle \tilde{L} | \hat{T}_{L}^\dagger \hat{T}_{L} | \tilde{L}\rangle, \\
&N_L^{\lambda T} = \langle \tilde{L} | \hat{\lambda}_{L}^\dagger \hat{T}_{L} | \tilde{L}\rangle.
\end{align}
\end{subequations}
The last term, $\mathbf{d}^{(2)}_{\mathrm{shift}}$, arises from the zeroth-order Hamiltonian $\boldsymbol{\mathcal{F}}$.
For example, the contributions from $\tilde{\Omega} \in \left\{ ab,T \right\}$ to $\mathbf{d}^{(2)}_\mathrm{shift}$ is
\begin{subequations}
\begin{align}
d^{(2)}_{\mathrm{shift},aa} & = -\sum_{b,T} \Lambda_{ab,T} \\
d^{(2)}_{\mathrm{shift},bb} & = -\sum_{a,T} \Lambda_{ab,T} \\
\bar{d}^{(2)}_{\mathrm{shift},rs} & = \sum_{L}\sum_{TU} \sum_{tu,vw} \bar{z}_{TU} V_{tv,L}^{T} \Gamma^{(3),LL}_{tu,vw,rs} V_{uw,L}^{U}.
\end{align}
\end{subequations}
The zeroth-order energy in the denominator is taken account by defining a norm-like quantity,
\begin{align}
& N^{\mathrm{shift}}_{L} = -\sum_{\tilde{\Omega}} \Lambda_{\tilde{\Omega}}^L, \\
& d^{(2)}_{\mathrm{shift},xy}  =
\left\{ \begin{array}{ll}
\displaystyle
\bar{d}^{(2)}_{\mathrm{shift},xy} - \sum_L N_L^{\mathrm{shift}} \langle \tilde{L} | \hat{E}_{xy} | \tilde{L} \rangle
& {x,y} \in {r,s} \\[8pt]
\displaystyle
\bar{d}^{(2)}_{\mathrm{shift},xy}
& \mathrm{otherwise}
\end{array} \right. \label{dshift}
\end{align}
Since $\Lambda$ and $\bar{z}$ involve both $\lambda$ and $T$ [Eq.~\eqref{LambdaOmega}], 
$\mathbf{d}^{(2)}_\mathrm{shift}$ is also a second-order contribution.
The correlated density matrices are then used to evaluate $Y_{xy}$,
as elaborated in Ref.~\onlinecite{Celani2003JCP}.

\begin{figure*}[tb]
	\includegraphics[width=0.8\linewidth]{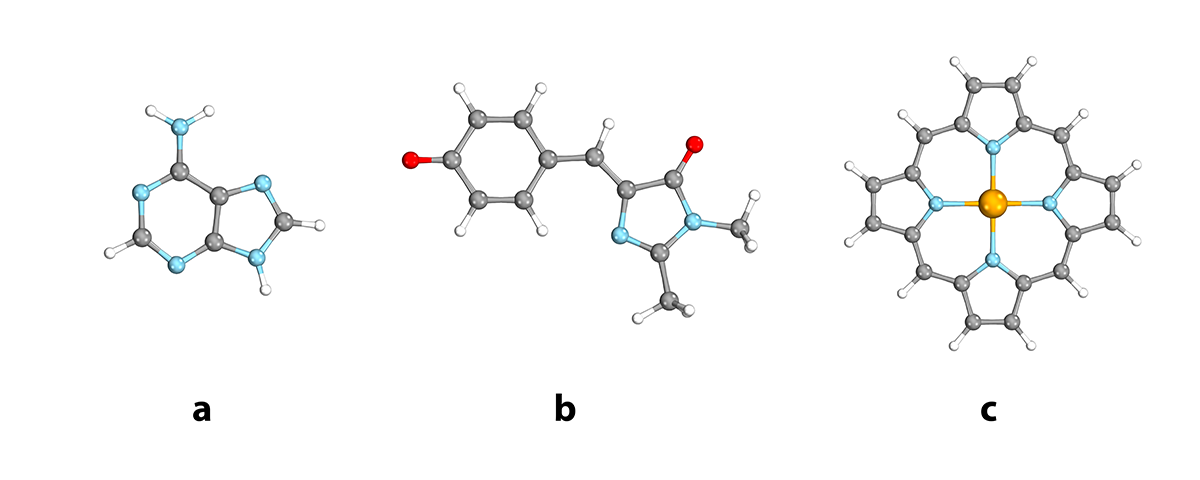}
	\caption{Optimized geometry of (a) adenine (b) $p-$HBDI$^-$ and (c) FeP (imaginary $\epsilon$ = 0.20 $E_\mathrm{h}$).
		Graphic created with IboView.\cite{Knizia2013JCTC,Knizia2015ACIE}
		\label{figure:01}}
\end{figure*}

Similarly, the CI derivatives can be divided into four components,
\begin{align}
\tilde{y}_{I,M} & = \frac{\partial \mathcal{L}_{\mathrm{imag},P}}{\partial \tilde{c}_{I,M}} \nonumber \\
& = \tilde{y}_{I,M}^\mathrm{(0)+(1)}+ \tilde{y}_{I,M}^{T\lambda} + \tilde{y}_{I,M}^{TT}  +
\tilde{y}_{I,M}^\mathrm{shift}\label{ytot}.
\end{align}
The counterpart in the CASPT2 nuclear gradient theory with the real shift is $\tilde{y}_{I,M} = \tilde{y}_{I,M}^\mathrm{(0)+(1)} + \tilde{y}_{I,M}^{T\lambda}$.\cite{Shiozaki2011JCP3,Vlaisavljevich2016JCTC}
The first two terms are
\begin{subequations}
\begin{align}
& \tilde{y}_{I,M}^\mathrm{(0)+(1)} = \sum_{N} R_{MP} R_{NP} \left( 2 \langle I | \hat{H} | \tilde{N}\rangle + \langle I | \hat{T}_{M}^\dagger \hat{H} | \tilde{N}\rangle + \langle \tilde{N} | \hat{T}_{M}^\dagger \hat{H} | I \rangle \right) \nonumber \\
& \quad\quad + \langle \tilde{M} | \hat{\lambda}^\dagger_{M} \hat{H} | I \rangle + \langle I | \hat{\lambda}^\dagger_{M} \hat{H} | \tilde{M} \rangle, \\
& \tilde{y}_{I,M}^{T\lambda} = \langle \tilde{M} | \hat{\lambda}_{M}^\dagger (\hat{H}^{(0)} - E_M^{(0)}) \hat{T}_{M} | I \rangle  + \langle I | \hat{\lambda}_{M}^\dagger (\hat{H}^{(0)} - E_M^{(0)}) \hat{T}_{M} | \tilde{M} \rangle \nonumber \\
& \quad\quad + 2 \sum_{rs} \langle I | \hat{E}_{rs} | \tilde{M} \rangle \left[ W_M \mathbf{g} ( \mathbf{d}^{(2)}_{T\lambda}) - N_M^{\lambda T} \mathbf{f} \right]_{rs},
\end{align}
\end{subequations}
and the additional terms are
\begin{subequations}
\begin{align}
& \tilde{y}_{I,M}^{TT} = R_{MP}^2 \left( \langle I | \hat{T}_{M}^\dagger \hat{H}|\tilde{M} \rangle + \langle \tilde{M}| \hat{T}_{M}^\dagger \hat{H} | I \rangle \right) \nonumber \\
& \quad\quad+ 2 R_{MP}^2 \langle I | \hat{T}_{M}^\dagger (\hat{H}^{(0)} - E_M^{(0)}) \hat{T}_{M} | \tilde{M} \rangle \nonumber \\
& \quad\quad + 2 \sum_{rs} \langle I | \hat{E}_{rs} | \tilde{M} \rangle \left[ W_M \mathbf{g} ( \mathbf{d}^{(2)}_{TT}) - N_M^{TT} \mathbf{f}  \right]_{rs}, \\
& \frac{1}{2} \tilde{y}_{I,M}^{\mathrm{shift}} = 
 \sum_{rs} \langle I | \hat{E}_{rs} | \tilde{M}\rangle \left[  W_M \mathbf{g} \left( \mathbf{d}^{(2)}_\mathrm{shift} \right) - N^\mathrm{shift}_M \mathbf{f} \right]_{rs} \nonumber \\
&\quad\quad + \sum_{\tilde{\Omega} \tilde{\Omega}^\prime} \tilde{\lambda}_{M,\tilde{\Omega}} \tilde{T}_{M,\tilde{\Omega}^\prime} \frac{\partial S_{\tilde{\Omega} \tilde{\Omega}^\prime}}{\partial \tilde{c}_{I,M}} \epsilon^2 
\Delta_{\tilde{\Omega}^\prime} \nonumber \\
&\quad\quad + \sum_{rs} e^{(1),MM}_{rs} \langle I | \hat{E}_{rs} | \tilde{M} \rangle \nonumber \\
&\quad\quad + \sum_{rs,tu} e^{(2),MM}_{rs,tu} \langle I | \hat{E}_{rs,tu} | \tilde{M} \rangle \nonumber \\
& \quad\quad + \sum_{rs,tu,vw} e^{(3),MM}_{rs,tu,vw} \langle I | \hat{E}_{rs,tu,vw} | \tilde{M} \rangle \nonumber \\
&\quad\quad + \sum_{rs,tu,vw} e^{(4),MM}_{rs,tu,vw} \sum_{xy} \langle I | \hat{E}_{rs,tu,vw,xy} | \tilde{M} \rangle f_{xy}. \label{yshift}
\end{align}
\end{subequations}
The density-like terms $\mathbf{e}^{(n)}$ are
\begin{align}
\mathbf{e}^{(n)} = \mathbf{e}^{(n)S} + \mathbf{e}^{(n)F},
\end{align}
where $\mathbf{e}^{(n)S}$ is defined in Eq.~\eqref{ens}, and $\mathbf{e}^{(n)F}$ is
\begin{align}
\mathbf{e}^{(n)F,MM} = \sum_{TU} \bar{z}_{TU} \frac{\partial }{\partial \mathbf{\Gamma}^{(n),MM}} \left( \mathbf{V}^\dagger \boldsymbol{\mathcal{F}} \mathbf{V} \right)_{TU}.
\end{align}
Note that $\mathbf{e}^{(n)F}$ does not appear in Eq.~\eqref{lagrangian},
as it also depends on the molecular integrals.
For example, the contribution from $\tilde{\Omega} \in \left\{ ab,T \right\}$ to $\mathbf{e}^{(3)F}$ is
\begin{align}
e^{(3)F,MM}_{rs,vw,tu} & = \sum_{TU} \bar{z}_{TU} V^T_{rv,M} V^U_{sw,M} f_{tu},
\end{align}
The working expressions for $\mathbf{d}^{(2)}$ and $\mathbf{e}$ in all other subspaces are compiled in the Supporting Information.
The Lagrange multipliers $w_{MN}$ and $z^c$ are then evaluated using the procedure described in the previous works\cite{Celani2003JCP,Shiozaki2011JCP3,Vlaisavljevich2016JCTC} as
\begin{subequations}
\begin{align}
w_{MN} & = -\frac{1}{2} \frac{1}{E_M^{(0)} - E_N^{(0)}} \sum_I \left( \tilde{c}_{I,M} \tilde{y}_{I,N} - \tilde{c}_{I,N} \tilde{y}_{I,M} \right), \\
z^c_{ij} & = - \frac{1}{2} \frac{Y_{ij} - Y_{ji}}{f_{ii}- f_{jj}}, \\
z^c_{ab} & = - \frac{1}{2} \frac{Y_{ab} - Y_{ba}}{f_{aa}- f_{bb}}.
\end{align}
\end{subequations}
Finally, the $Z$-vector equation is solved using $Y$ and $y$ as the source terms.

\begin{figure}
\includegraphics[width=\linewidth]{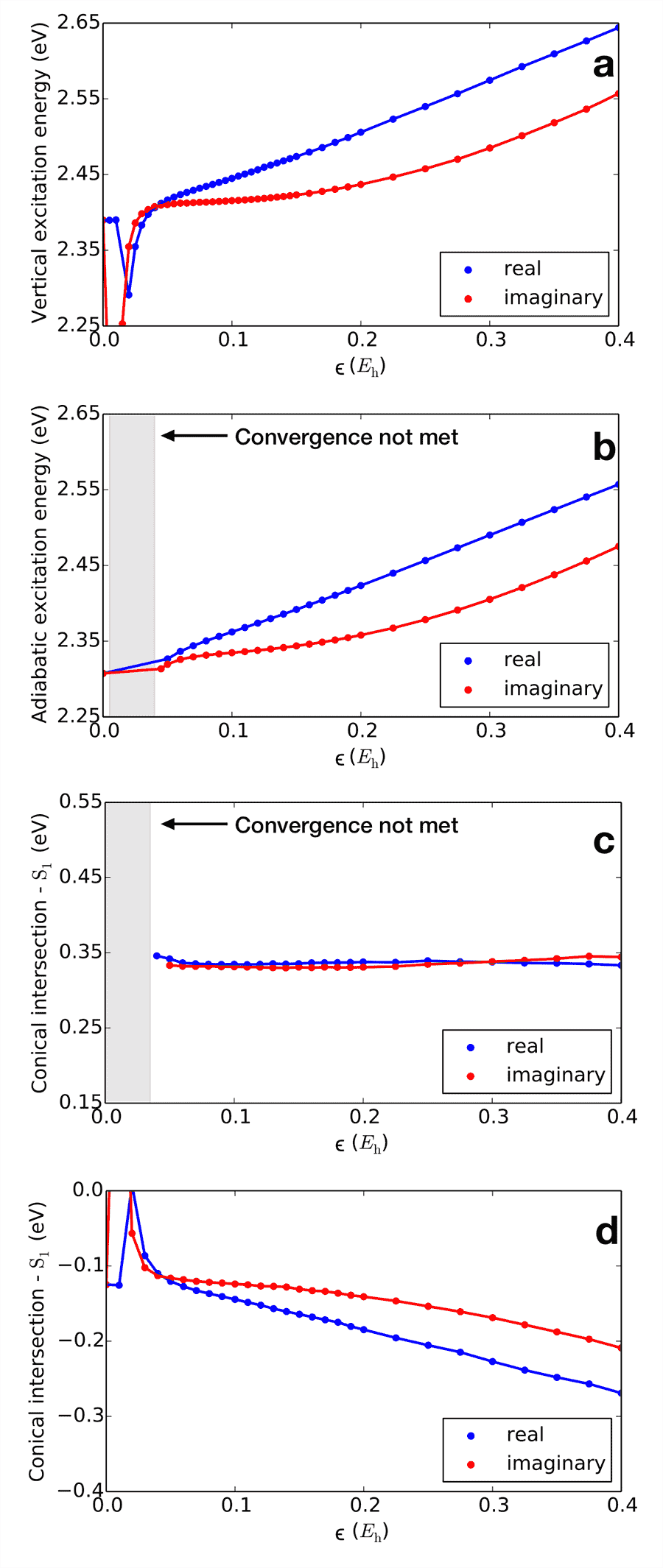}
\caption{Excitation energies (eV) for $p-$HBDI$^-$. Plot of
 (a) vertical excitation energy of $\mathrm{S}_1$,
 (b) adiabatic excitation energy of $\mathrm{S}_1$--$\mathrm{S}_0$ and
 (c) the difference in energy between the conical intersection and the Franck--Condon point for the $P$ conformer and
 (d) that for the $I$ conformer. \label{figure:02}}
\end{figure}

\section{Numerical Examples}

The numerical results for the imaginary shift formalism are presented in the following subsections.
Geometry optimizations were performed using XMS-CASPT2 as implemented in the {\sc bagel}
program with both real and imaginary shifts for comparison.
Calculations on adenine and the deprotonated form of 4-hydroxybenzylidene-1,2-dimethylimidazolinone ($p-$HBDI$^-$) were performed with cc-pVDZ\cite{Dunning1989JCP} and the corresponding density-fitting basis set.
SVP\cite{Schafer1992JCP} and the associated fitting basis was used for iron (II) porphyrin (FeP).

Calculations performed on $p-$HBDI$^-$ were performed with an active space consisting of four electrons in three orbitals (4\textit{e}, 3\textit{o}).
We used a minimal active spaces of (4\textit{e}, 4\textit{o}) for adenine. 
The inorganic porphyrin complex FeP was optimized for the low spin singlet state with scalar relativistic effects using the Douglas--Kroll--Hess Hamiltonian.\cite{Douglas1974,Hess1986PRA}
We employed an active space of (10\textit{e}, 9\textit{o}) for FeP, which is a minimal active space for metal porphyrin--ligand binding,\cite{Jensen2005JInorgBiochem,Falahati2018NatComm}
which includes five metal 3$\textit{d}$ orbitals and four Gouterman orbitals on the ligand.\cite{Gouterman1959JCP,Gouterman1961JMS}
The optimized structures for all of the molecules in this study are depicted in Fig.~\ref{figure:01}.

\subsection{Accuracy}

\begin{figure}[t]
\includegraphics[width=\linewidth]{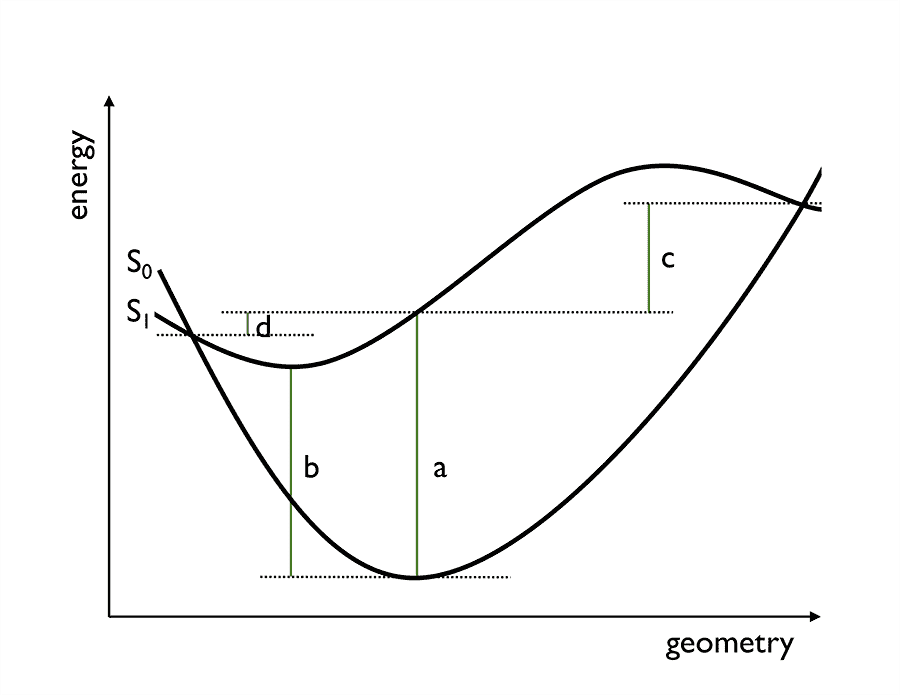}
\caption{Pictorial representation of $\mathrm{S}_0$ and $\mathrm{S}_1$ surfaces.
a, b, c, and d corresponds to the panels in Fig.~\ref{figure:02}.
\label{figure:03}}
\end{figure}

$p-$HBDI$^-$ is an anionic green fluorescent protein model chromophore (Fig.~\ref{figure:01}).
We computed the vertical excitation energy at the S$_0$ geometry, the adiabatic excitation energy, and the energies at the conical intersection points relative to the Frank--Condon point with various values of real and imaginary shifts.
The results are shown in Fig.~\ref{figure:02}. The schematic representation of each of the quantities is presented in Fig.~\ref{figure:03}.

The $\mathrm{S_1}$ vertical excitation energies of $p-$HBDI$^-$ are shown in Fig.~\ref{figure:02}(a) as a function of shift value.
The computed energies increase with increasing shift value, because less correlation is included in the calculation at larger shift values.\cite{Roos1996JMS}
However the results obtained with the imaginary shift are less sensitive to variation of the shift parameters than those with the real shift. 
This must be ascribed to the quartic behavior of the error in the imaginary shift approach whereas the asymptotic behavior with the real shift has a linear dependence on the shift parameter [see Eq.~\eqref{realaa} and \eqref{imagaa}].
For very small shift values near zero ($\textless$ 0.04 $E_\mathrm{h}$), the presence of intruder states is apparent.

By fitting the data in the range of $\epsilon=0.05-0.20~E_\mathrm{h}$, the extrapolated excitation energy is found to be 2.40 (2.39) eV using the imaginary (and real) shift.
If one chooses a small shift parameter, for example, $\epsilon= 0.05~E_\mathrm{h}$, the results from real (2.42 eV) and imaginary (2.41 eV) shift calculations quantitatively match with each other.
However, with a practical values of $\epsilon$ that are commonly used to avoid the intruder state problem,\cite{Roos1996JMS}
for instance, $\epsilon$ = 0.20 $E_\mathrm{h}$, the vertical excitation energy is computed to be 2.51 and 2.44 eV with the real and imaginary shifts, respectively.
Given that the expected value is $\approx$ 2.40 eV, the relative error with the imaginary shift (40 meV) is less than half of that with the real shift (110 meV).
The vertical excitation energy computed with $\epsilon$ = 0.40 $E_\mathrm{h}$ gives 2.64 eV and 2.56 eV for the real and
imaginary shifts, respectively.
At this value of $\epsilon$, the error due to the real shift is comparable to the intrinsic accuracy of the CASPT2 model.\cite{Schreiber2008JCP} 

The adiabatic excitation energies [Fig.~\ref{figure:02}(b)], based on the geometry optimization of both $\mathrm{S}_0$ and $\mathrm{S}_1$ states similarly show linear and quartic for the real and imaginary shifts, respectively.
For calculations with the shift values between 0.005 and 0.040 $E_\mathrm{h}$, convergence was not met due to an intruder in the $\mathrm{S}_2$ state during the geometry optimization on the $\mathrm{S}_1$ surface.
The extrapolated adiabatic excitation energy using the same procedure as above with the real and imaginary shift results is found to be 2.31 and 2.30 eV.
With $\epsilon = 0.20~E_\mathrm{h}$ that is commonly used in practical calculations, the adiabatic excitation energy is 2.42 and 2.36 eV for the real and imaginary shifts, respectively,
which means that the error with the imaginary shift (60 meV) is roughly half the error with the real shift (120 meV) calculation.
The same trend holds for larger values of $\epsilon$.

Figures~\ref{figure:02}(c) and (d) show the energies at the conical intersections between the $\mathrm{S}_1$ and $\mathrm{S}_0$ surfaces 
relative to the $\mathrm{S}_1$ energy at the Franck--Condon point. We considered both the phenoxy ($P$)
and imidazolinone ($I$) twisted conformers of $p-$HBDI$^-$.
For the $P$ conformer, the resulting energies remain constant with respect to the shift parameters owing to fortuitous error cancellation, in both imaginary and real shift cases.
The sensitivity of the result is, however, apparent for the $I$ conformer [Fig.~\ref{figure:02}(d)];
for the small values of $\epsilon$ between 0.05--0.20 $E_\mathrm{h}$, the results computed with imaginary shifts are nearly constant, whereas with the real shift the results decrease linearly.
For larger shift values above 0.20 $E_\mathrm{h}$, the results are diverging at the nearly the same rate.
At $\epsilon = 0.20~E_\mathrm{h}$, the results with the real and imaginary shifts are $-0.18$ and $-0.14$~eV, which are to be compared with the extrapolated 
values $-0.10$ and $-0.11$ eV.

\begin{table}[t]
\caption{Root-mean-square deviation ($\AA$) of $p-$HBDI$^-$ $\mathrm{S}_0$ geometry relative to that computed with imaginary $\epsilon$ = 0.20 $E_\mathrm{h}$.  \label{table:01}}
\begin{ruledtabular}
\begin{tabular}{ccc}
$\epsilon$ ($E_\mathrm{h}$) & Real    & Imaginary \\
\hline                                                                                     
0.00                        & 0.00135 & 0.00139   \\
0.01                        & 0.00135 & 0.00754   \\
0.10                        & 0.00054 & 0.00051   \\
0.20                        & 0.00132 & ------    \\
0.30                        & 0.00278 & 0.00028   \\
0.40                        & 0.00432 & 0.00078   \\ 
0.60                        & 0.00770 & 0.00278   \\ 
0.80                        & 0.01102 & 0.00584   \\ 
1.00                        & 0.01418 & 0.01003   \\ 
\end{tabular}
\end{ruledtabular}
\end{table}
\begin{table*}[tb]
\caption{Wall times in seconds for representative steps in XMS-CASPT2 nuclear gradient evaluation. The timing was measured using 16 nodes of a SandyBridge cluster purchased in 2012 (Xeon E5-2650 2.00GHz, total of 256 CPU cores).  \label{table:02}}
\begin{ruledtabular}
\begin{tabular}{cccccrrrrrrr}
System             & Atoms / Electrons & Basis\footnotemark[1] & CAS    & States & \multicolumn{1}{c}{Amp.\footnotemark[2]} & \multicolumn{1}{c}{$\lambda$\footnotemark[3]} & \multicolumn{1}{c}{Den. (shift)\footnotemark[4]} & \multicolumn{1}{c}{CI deriv.} & \multicolumn{1}{c}{$Z$ vector} &\multicolumn{1}{c}{Total\footnotemark[5]} \\
\hline                                                                                                                                                                      
\multicolumn{11}{c}{Real shift}\\
adenine            & 15  /  70        & 165 (815)             & (4\textit{e}, 4\textit{o})  & 5   & 20  & 18  & 6.2 (--)  & 19   & 4.6 & 84   \\
$p-$HBDI$^-$       & 27  / 114        & 279 (1373)            & (4\textit{e}, 3\textit{o})  & 3   & 59  & 49  & 25 (--)   & 74   & 14  & 257  \\
FeP                & 37  / 186        & 427 (2288)            & (10\textit{e}, 9\textit{o}) & 5   & 744 & 325 & 297 (--)  & 879  & 160 & 2947 \\
\multicolumn{11}{c}{Imaginary shift}\\
adenine            & 15  / 70         & 165 (815)             & (4\textit{e}, 4\textit{o})  & 5   & 22  & 21  & 7.4 (0.8) & 20   & 4.5 & 87   \\
$p-$HBDI$^-$       & 27  / 114        & 279 (1373)            & (4\textit{e}, 3\textit{o})  & 3   & 61  & 56  & 26 (0.8)  & 75   & 14  & 260  \\
FeP                & 37  / 186        & 427 (2288)            & (10\textit{e}, 9\textit{o}) & 5   & 700 & 336 & 729 (383) & 1028 & 196 & 3387 \\
\end{tabular}
\end{ruledtabular}
\footnotetext[1]{The number of basis functions. The numbers in parentheses are the number of auxiliary functions.}
\footnotetext[2]{Total time for the CASPT2 amplitude equation, see Sec.~\ref{energysec}.}
\footnotetext[3]{Total time for the CASPT2 Lambda equation [Eq.~\eqref{Lambda_now}].}
\footnotetext[4]{Time for computing the correlated density matrices (which includes timing for $\tilde{y}^\mathrm{shift}$). The numbers in parentheses are the time for computing $\mathbf{d}^{(2)}_{\mathrm{shift}}$ and $\tilde{y}^\mathrm{shift}_{I,M}$ [Eqs.~\eqref{dshift}~and~\eqref{yshift}], which are unique to the CASPT2 nuclear gradients with the imaginary shift.}
\footnotetext[5]{Total wall time for a geometry optimization step, which includes the time for CASSCF and CASPT2 energy evaluation and computation of MO integrals and reference RDMs.}
\end{table*}

A similar trend in error is observed for the optimized geometries.
To illustrate, Table \ref{table:01} lists the root-mean-square deviation (RMSD) in $\AA$ngstr{\"o}m for the $\mathrm{S}_0$ geometry computed at various shift parameters. We used the geometry calculated with the imaginary shift $\epsilon$ = 0.20 $E_\mathrm{h}$ as a reference.
The structural differences with the real or imaginary shift, of various values, are under a hundredth of an $\AA$ngstr{\"o}m except for when $\epsilon$ is taken to be as large as 1.00 $E_\mathrm{h}$.
The RMSD tends to increase with increasing shift values, but more slowly with imaginary shift.

\subsection{Timing}

The computational cost of geometry optimization with imaginary shift was assessed for adenine, $p-$HBDI$^-$, and FeP and compared against that with the real shift.
To make the comparison consistent, all of the calculations were performed using 16 nodes of a Xeon E5-2650 cluster (SandyBridge 2.0~GHz, 32 CPUs/256 CPU cores, purchased in 2012). 
All of the timing calculations were performed using $\epsilon$ = 0.20 $E_\mathrm{h}$.
Our implementation does not exploit spatial symmetry of molecules.
The results are compiled in Table \ref{table:02}.

When the active spaces are small, the wall times for calculating the CASPT2 nuclear gradients with real and imaginary shifts were found to be essentially identical.
For instance, one geometry optimization step for adenine with CAS(4$e$, 4$o$) took 84 and 87 seconds, respectively, using the real and imaginary shifts.
The same held for the geometry optimization of $p$--HBDI$^-$ with CAS(4$e$, 3$o$), which took 257 and 260 seconds, respectively. 
Of these timings, roughly 20--25\% of time was spent for CASPT2 energy evaluation, 20\% for solving the $\lambda$-equation, and 25--30\% for computing the CI derivatives.
The rest was due to the computation of correlated density matrices and solution of the $Z$-vector equation. 

When a large active space was used, however, the difference in the computational costs became noticeable, though the difference was still minor. 
For example, a geometry optimization step for FeP with CAS(10$e$, 9$o$) took 2947 and 3387 seconds with the real and imaginary shifts, respectively,
indicating that the nuclear gradient evaluation with imaginary shift for this case was 15\% more expensive than the real shift counterpart.

The timing difference between the real and imaginary shift cases is mainly ascribed to the computational cost required for evaluating the imaginary shift terms, $\mathbf{d}^{(2)}_\mathrm{shift}$ in Eq.~\eqref{d2tot} and $\tilde{y}_{I,M}^\mathrm{shift}$ in Eq.~\eqref{ytot},
which roughly scales $O(N_\mathrm{act}^9)$ (see the Supporting Information).
Therefore, as the number of the active orbitals increases, the additional cost of evaluating these terms is expected to be more pronounced.
For FeP, the wall time for computing the correlated density matrices with the imaginary shift was 729 seconds,
among which the imaginary shift term was responsible for 383 seconds (52\%). 
This is in contrast to the adenine and $p-$HBDI$^-$ cases where the times of computing $\mathbf{d}^{(2)}_\mathrm{shift}$ and $\tilde{y}_{I,M}^\mathrm{shift}$ were less than a second, constituting only a fraction of the time for computing the correlated density matrices.

Compared to the $\mathbf{d}^{(2)}_\mathrm{shift}$ and $\tilde{y}_{I,M}^\mathrm{shift}$, the computational cost for the other additional terms was found only marginal.
The last term in Eq.~\eqref{Lambda_now} requires only one additional evaluation of a residual-like term at the beginning of the $\lambda$-iteration.
As a consequence, the total time for the $\lambda$-equation only slightly increased with imaginary shift, compared to that with real shift, by
3, 7, and 11 seconds for adenine, $p-$HBDI$^-$, and FeP, respectively.
Evaluation of yet another additional terms, $\mathbf{d}^{(2)}_{TT}$ in Eq.~\eqref{d2tot} and $\tilde{y}_{I,M}^{TT}$ in Eq.~\eqref{ytot}, can be combined with the conventional terms; therefore, the increase in the computational cost due to these terms was
found to be not significant either; they took 0.4, 1, and 49 seconds for adenine, $p-$HBDI$^-$, and FeP, respectively. 

\section{Conclusions}

We have derived and implemented the nuclear gradients for CASPT2 with the imaginary shift by extending the CASPT2 nuclear gradient code for the real shift.
The numerical results for the vertical and adiabatic excitation energies and the energy differences between the conical intersections and the Frank--Condon point for $p-$HBDI$^-$ showed that
the results were less sensitive to variation of imaginary shift values compared to those with real shift.
When small active spaces were used, the additional cost for computing CASPT2 nuclear gradients with imaginary shift was found to be marginal.
In a calculation for FeP for which a larger active space [CAS(10$e$, 9$o$)] was used, we observed that the wall times with imaginary shift were roughly 15\% more than that with real shift. 
The difference has been shown to be due to the computation of correlated density-like quantities that are associated with the imaginary shift terms.
The programs have been interfaced to the {\sc bagel} package, which is publicly available for use in chemical applications.

\section{Acknowledgments}
This work has been in part supported by National Science Foundation [ACI-1550481 (JWP) and CHE-1351598 (TS)].
RA-S has been supported by Air Force Office of Scientific Research (AFOSR FA9550-18-1-0252).
JWP has also been supported by the National Research Foundation of Korea (NRF) grant funded by the Korea government (MSIT) (No. 2019R1C1C1003657).

\section{Supporting Information}
Complete set of equations, including working expressions for $\mathbf{d}^{(2)}$ and $\mathbf{e}$, and tables that compile the raw data for Fig.~\ref{figure:02}.

\end{document}